\documentclass[article,11pt,leqno]{article}
\usepackage{Other/Templates/paper}

\newcommand{\titleLong}{\vspace{-2.5cm}The Impact of School and Family Networks on COVID-19 Infections Among Dutch Students: A Study Using Population-Level Registry Data}
\hypersetup{pdftitle={\titleLong}}

\usepackage{soul} % Highlighted todo tasks
\begin{document}

%% TITLE PAGE including abstract
\title{\titleLong}

% Order to be determined depending on who does what
\author[1, 2, *]{Javier Garcia-Bernardo}
\author[3]{Christine Hedde-von Westernhagen}
\author[4]{Tom Emery}
\author[5]{Albert Jan van Hoek}
\affil[1]{ODISSEI Social Data Science (SoDa) Team \& Department of Methodology and Statistics, Utrecht University, Utrecht, the Netherlands}
\affil[2]{Centre for Complex Systems Studies, Utrecht University, Utrecht, the Netherlands}
\affil[3]{Department of Industrial Engineering and Innovation Sciences, Eindhoven University of Technology, Eindhoven, the Netherlands}
\affil[4]{ODISSEI \& Department of Public Administration and Sociology, Erasmus University, Rotterdam, the Netherlands}
\affil[5]{Centre for Infectious Diseases Control, National Institute for Public Health and the Environment, Bilthoven, the Netherlands}

\date{}%Version: \today}                       

\begin{titlepage}
\maketitle

\vspace{-0.7cm}
\begin{abstract}
Understanding the impact of different types of social interactions is key to improving epidemic models.
Here, we use extensive registry data---including PCR test results and population-level networks---to investigate the impact of school, family, and other social contacts on SARS-CoV-2 transmission in the Netherlands (June 2020--October 2021). 
We isolate and compare different contexts of potential SARS-CoV-2 transmission by matching pairs of students based on their attendance at the same or different primary school (in 2020) and secondary school (in 2021) and their geographic proximity. We then calculate the probability of temporally associated infections---i.e. the probability of both students testing positive within a 14-day period.

Our results highlight the relative importance of household and family transmission in the spread of SARS-CoV-2 compared to school settings. The probability of temporally associated infections for siblings and parent-child pairs living in the same household ranged from 22.6--23.2\%. Interestingly, a high probability (4.7--7.9\%) was found even when family members lived in different households, underscoring the persistent risk of transmission within family networks. In contrast, the probability of temporally associated infections was 0.52\% for pairs of students living nearby but not attending the same primary or secondary school, 0.66\% for pairs attending different secondary schools but having attended the same primary school, and 1.65\% for pairs attending the same secondary school. It is worth noting, however, that even small increases in school-related infection probabilities can trigger large-scale outbreaks due to the dense network of interactions in these settings. Finally, we used multilevel regression analyses to examine how individual, school, and geographic factors contribute to transmission risk. We found that the largest differences in transmission probabilities were due to unobserved individual (60\%) and school-level (35\%) factors. 
Only a small proportion (3\%) could be attributed to geographic proximity of students or to school size, denomination, or the median income of the school area.

%Our insights into the transmission dynamics of SARS-CoV-2 within Dutch educational institutions could improve the future application of network models in public health.
\end{abstract}

\textit{\vspace{-0.5cm}Keywords}: COVID-19; transmission dynamics; school networks; family networks; registry data
    
\end{titlepage}

%% INTRODUCTION
\section{Introduction}\label{s:introduction}
%1\. Context/relevance
Epidemic models that explicitly incorporate network structure have gained traction during the COVID-19 pandemic \parencite{bustamante-castanedaEpidemicModelNetwork2021, prasseNetworkinferencebasedPredictionCOVID192020, firthUsingRealworldNetwork2020, sanchezMultilayerNetworkModel2022, cui2021network}.
Including information about the intricate network structure of the population allows for better predictions of the shape of the epidemic curve, the regions or population groups likely to be infected given observed cases, and the types of contacts relevant for transmission. All three are highly relevant for public health responses. 

%2\. Research gap
Despite advances in network-based epidemic modeling during the COVID-19 pandemic, a significant gap remains in measuring the relative impact of different types of school and family contacts on transmission dynamics. Previous studies on influenza \parencite{endoClassroomTransmissionPatterns2021, cauchemezRoleSocialNetworks2011}, and also SARS-CoV-2 \parencite{vanierselEmpiricalEvidenceTransmission2023, cmmidcovid-19workinggroupImplicationsSchoolhouseholdNetwork2021c} highlight the importance of schools as bridges for disease transmission between households, schools and other social contact areas.
Indeed, in an attempt to contain the spread of the virus, governments around the world closed schools, resulting in large learning losses, especially among students from less-educated families \parencite{engzell2021learning}.

At the same time, studies in diverse countries such as the United Kingdom, Australia, and Singapore emphasize that transmission within schools can be managed with interventions such as physical distancing, air filtering and rapid isolation, and that household contacts remain a more prominent pathway for transmission \parencite{thompson2021staff, macartney2020transmission, cordery2022transmission, heavey2020no, ismail2021sars, yung2021novel}. For instance, \textcite{cordery2022transmission} observed minimal school-based transmission when precautions were in place, contrasting with high secondary transmission rates in households, likely due to prolonged close contact and viral shedding. Furthermore, in Wales, \textcite{thompson2021staff} found that while students faced increased risks of infection from peers in their immediate year groups, the total number of cases in a school was not associated with an increased risk for staff or pupils. Similarly, \textcite{macartney2020transmission} observed low SARS-CoV-2 transmission rates in Australian educational settings, suggesting that schools did not contribute significantly to COVID-19 spread when effective case-contact testing and epidemic management strategies were in place.

%3\. Specific issue
Our study adds to this body research by examining the impact of family and school contacts on COVID-19 transmission among Dutch students using detailed population-level registry data from the Netherlands. We specifically examine students who transitioned from primary to secondary school in 2021, which takes place at age 12.  Focusing on this transition allows us to distinguish whether infections occur primarily at school, through social ties inherited from primary school, or through non-school interactions such as community transmission.  Specifically, we match pairs of students who attended primary school together in 2020 and either attended separate secondary schools or the same school in 2021. Given the large student segregation at schools \parencite{kazmina2024socio}, we expect students who attended the same primary school to be similar to each other. We compare these groups of students to a reference group of students who did not attend primary or secondary school together. We then calculate the probability of temporally associated infections for the different groups as a function of the distance between the students' homes. Next, we compare our results to the probabilities of temporally associated infections among different family members (siblings, parent-child, co-parents) living in the same house or at varying distances. Finally, we run a series of multilevel regressions to understand the heterogeneity between schools.

%5\. Provide a succinct preview of the main findings or results, setting the stage for the subsequent sections of the paper.
Our findings show that family ties contribute strongly to the spread of SARS-CoV-2. While attending the same school increased the probability of temporally associated infections from 0.5\% to 1.6\%, the probability of associated infections was much higher for family members living in the same house (25--50\%) and even for family members living at different addresses (around 10\%). During the period studied in this paper, temporally associated infections in primary schools were rare. 
These results align with previous literature showing a high frequency of secondary transmission in households \parencite{cordery2022transmission, thompson2021staff} and low frequency of transmission in schools.
%It is worth noting that even small increases in the probability of co-infection in schools can lead to large outbreaks. This is because students have so many more interactions at school, and the probability is compounded. 
Examining heterogeneity at the school level, we found that factors such as the distance between the students' homes, school size, the median income of the postcode area of the school, and school denomination explained only 3\% of the variance in outcomes. Most of the variance manifested at the individual level (60\%) and at the school level (35\%). 

The paper proceeds as follows: Section \ref{s:methods} details the data, the matching of students and the analysis of the data. Section \ref{s:results} details the probability of temporally associated infections for different subgroups. Section \ref{s:conclusion} concludes and discusses the potential of administrative data for epidemic studies.

%% METHODS
\section{Data and Methods}\label{s:methods}

\subsection{Main Datasets and Network Construction}
Our analysis integrates two main datasets from Statistics Netherlands (CBS): the COVID-19 PCR-test data and the population network data. Every CBS dataset can be linked to each other at the individual-level through a unique identifier. We provide below a short summary of the data processing steps. A detailed explanation of all datasets and variables can be found in the online Supplementary Information.

\begin{figure}[h!]
\includegraphics[width=0.95\textwidth]{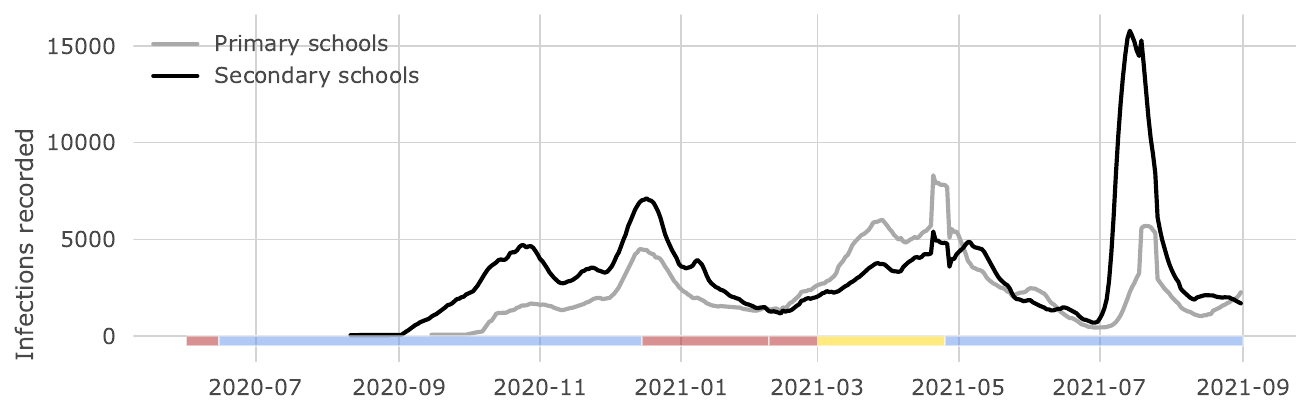}
    \caption{\small \textbf{Number of SARS-CoV-2 infections}. Infections are measured by the municipal health services using PCR-tests, and displayed for students attending primary (gray line) and secondary schools (black lines), aggregated per week (using a rolling window) over the time studied. To preserve the privacy of those individuals and in line with CBS regulations, only weeks with at least 10 cases are shown. School closures are shown at the bottom of the figure. Time periods where secondary schools were closed are marked in red: June 1st--15th (2020) and December 15th (2020)--March 1st (2021). Time periods where secondary schools were open are marked in yellow (open with restrictions, March 1st--April 26th (2021) and blue (open without restrictions). Primary schools were open from February 8th--March 1st (2021). }
    \label{fig:closures}
\end{figure}

The first dataset is the \textit{COVID-19 test} dataset, which includes all PCR-tests conducted by municipal health services in the Netherlands outside of a hospital setting between June 2020 and September 2021. Schools were open for the majority of the period studied (Fig.~\ref{fig:closures}). Since reinfections were unusual for the period studied, we retained for each person the first recorded infection.  

The second dataset is the \textit{Person Network} dataset, which contains formal relationships---i.e., family, school and household relationships recorded officially by the government---between individuals connecting the entire population of the Netherlands. These connections indicate ``a highly increased probability that two individuals interact socially'' \parencite{vanderlaanWholePopulationNetwork2022}. 
Administrative networks bear a novel opportunity to researchers studying social processes since they do not suffer from common drawbacks of studies based on surveys, digital trace data, or contact tracing such as non-response bias, selection bias, or social desirability effects \parencite{vanderlaanWholePopulationNetwork2022}. Furthermore, these data are readily available in the Netherlands as well as many other countries, and could lower the burden of additional data collection efforts to inform policy decisions in a pandemic situation.

We constructed school networks using educational records from primary and secondary schools. Educational records connect students to their schools, year of education, and program tracks. 
We included students who did not attend special schools---schools servicing students with special needs such as blind or deaf individuals, where infection dynamics are likely to be different. To compare infections arising from school, family and non-school interactions (Section \ref{s:m:matching}--\ref{s:m:tai}) we focus on students transitioning to secondary school in September 2021. For the multilevel regression analysis (Section \ref{s:m:regression}) we focus on students registered in primary schools. This approach allows us to compare transmission dynamics across distinct social environments by examining both primary and transitioning secondary students.

To analyze the role of family ties in the transmission of SARS-CoV-2, we collected all family pairs in the following categories: Full-siblings, co-parents (two adults being the parents of the same child) and parent-child. We extracted family networks using data derived from parent-child records \parencite{vanderlaanWholePopulationNetwork2022}. For example, siblings are recorded if they share at least a parent or if their parents are partners. Different types of siblings---such as half-siblings (who share one biological parent) and step-siblings (who have no biological parents in common and are related through their parents’ relationship)---may have very different levels of closeness. Some might grow up together, some might grow up in different houses (especially those who share the same father), while others might become siblings as adults and have less frequent contact. To better estimate the probability of temporally associated co-infections in siblings that are likely to keep regular contact, we focus only on full-siblings.

Finally, for the multilevel regression analyses we classified schools according to their denomination. The school denomination denotes the type of school and is correlated with attitudes towards COVID-19. In the Netherlands, parents have the right to choose schools that match their values. A majority of schools are Christian (either Protestant, Catholic, Evangelic or Reformist), while around one third are public schools \parencite{engzell2021learning}. Other denominations include Islamic schools and Anthroposophic. We also included in the regression analyses the school size, the median income of the school's neighborhood (at the 4-digit postcode level, an administrative area equivalent to a neighborhood with an average population size of 4,314 and a maximum population of 28,190 \parencite{statistiekKerncijfersPostcode} and the distance between the house address of each pair of students.  For privacy considerations, the location of the houses is only known at a resolution of 100x100 m\textsuperscript{2}. We assigned all individuals to the last known address before 2021 and kept individuals who remained living in the Netherlands throughout 2021. We estimated the distance between two households as the euclidean distance plus 52 meters---the average distance between two random points in a 100x100 m\textsuperscript{2} square. This implies that students living in the same 100x100 m\textsuperscript{2} are estimated to live at a distance of 52 meters. Students living in the same household (sharing the same house ID) were set at a distance of 0 meters.

\subsection{Matching Students in Groups of Increasing Level of Contact} \label{s:m:matching}
\begin{figure}[h!]
    \includegraphics[width=0.7\textwidth]{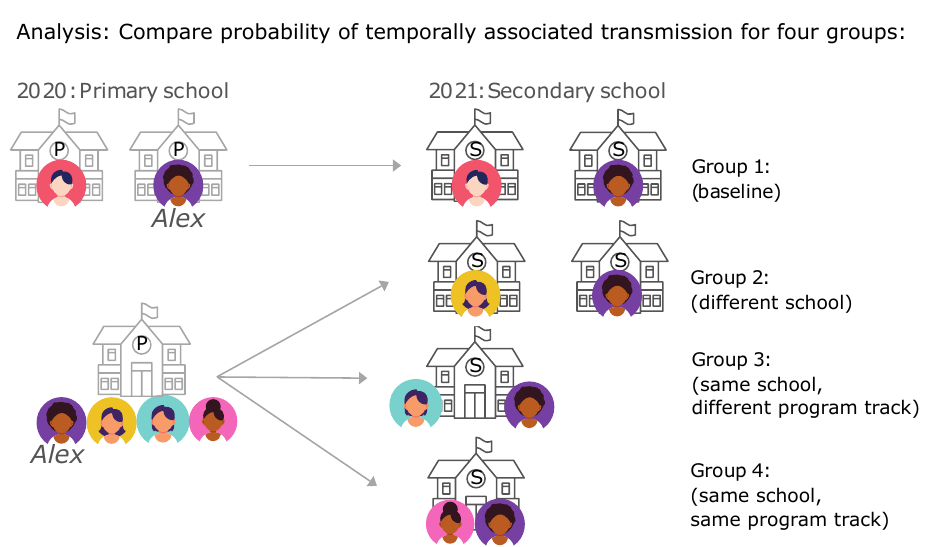}
    \caption{\small \textbf{Illustration of the different types of student pairs}, with increasing level of expected contact. Each student in the data is paired with different type of students. The example shows the process for one student named Alex. Group 1: Students who attend a different primary school than Alex and a different secondary school. Group 2: Students who attend the same primary school than Alex and a different secondary school. Group 3: Students who attend the same primary and secondary school than Alex, but are placed in a different program track in the secondary school. Group 4: Students who attend the same primary and secondary school than Alex, and are placed in the same program track in the secondary school. }
    \label{fig:design_groups}
\end{figure}

We first analyzed the role of schools in the transmission of SARS-CoV-2 by focusing on students transitioning from primary school to secondary school in 2021 (illustrated in Fig.~\ref{fig:design_groups}). Focusing on this transition allows us to distinguish whether infections mainly occur at school, or from non-school interactions such as community transmission.  We create four groups of student pairs representing increasing level of contact:
\textit{Group 1 (Baseline)}: Pairs of students who did not attend the same primary or secondary school. Since we are interested in a comparison group of pairs of students living near each other, we oversampled pairs of students living within the same municipality. 
\textit{Group 2 (Same background)}: Pairs of students who attended the same primary school (and will have a similar social background) but not the same secondary school. \textit{Group 3 (Same school, different program track)}: Pairs of students who attended both primary and secondary schools together but were not in the same program track in secondary school---i.e, they attend different classrooms. \textit{Group 4 (Same school, same program track)}: Pairs of students who attended both primary and secondary schools together and were in the same program track in secondary school.

Finally, we created three separate categories for twins, which we identify as pairs of students living in the same household and attending the same school year. \textit{Twins 2 (Same background)}: Pairs of twins who attended the same primary school but not the same secondary school. \textit{Twins 3 (Same school, different program track)}: Pairs of twins who attended both primary and secondary schools together but were not in the same program track in secondary school---i.e, they will attend different classrooms. \textit{Twins 4 (Same school, same program track)}: Pairs of twins who attended both primary and secondary schools together and were in the same program track in secondary school. Unsurprisingly, we were unable to create a group \textit{Twins 1 (Baseline)} since there were less than 10 twin pairs in the studied cohort which did not attend the same primary or secondary school together.

\subsection{Probability of Temporally Associated Infection}\label{s:m:tai}

We calculated the probability of temporally associated infections within a 14-day period for each group as a function of distance between the places of residence of both students of family members. To preserve the individuals' privacy, results can only be exported from the secure computer of CBS in groups of at least 10 individuals. Because of this, we calculated the number of student or family pairs and number of temporally associated infected pairs for the following distance bins: 0m, 0-300m, 300-1000m, 1000-3000m, 3000-10,000m, 30,000+. Each bin excludes its left boundary (e.g., 0–300m includes distances greater than 0m but up to 300m), except for the 0m bin, which represents individuals living in the same household.  The 14-day period was chosen based on the approximately 7 days incubation and generation periods for SARS-CoV-2 \parencite{zhaoEstimatingGenerationInterval2021}, but the results are robust to changes of this threshold.

The probability of temporally associated infections within a distance bin $d$ is calculated as $P_{\text{temp}}(d) = \frac{\sum_{i,j} I_{ij}(d)}{\sum_{i,j} N_{ij}(d)},$ where $I_{ij}(d)$ equals one if the student or family pair $(i,j)$ living within the distance bin $d$ tested positive within a 14-day period, and $N_{ij}(d)$ is the total number of student or family pairs within the distance bin $d$.

The number of student and family pairs ($\sum_{i,j} N_{ij}(d)$) and associated temporally associated infections ($\sum_{i,j} I_{ij}(d)$) is given in Table~\ref{t:summary_pairs} and in Table~\ref{t:summary_pairs_family} respectively for school and family ties.

\begin{table}[ht]
\small
\begin{tabular}{r|llllllll}
Group & \multicolumn{2}{l}{G1 (baseline)} & \multicolumn{2}{l}{G2 (same background)} & \multicolumn{2}{l}{G3 (same school)} & \multicolumn{2}{l}{G4 (same program)} \\
 & N & N\_inf & N & N\_inf & N & N\_inf & N & N\_inf \\
\midrule
Same house & <10 & <10 & 84 & 28 & 62 & 24 & 107 & 54 \\
<0.3 km & 908 & 11 & 18,726 & 141 & 4,886 & 59 & 4,045 & 71 \\
0.3-1 km & 6,856 & 40 & 80,741 & 533 & 19,883 & 222 & 16,301 & 257 \\
1-3 km & 33,199 & 150 & 78,547 & 501 & 18,708 & 221 & 14,768 & 254 \\
3-10 km & 105,180 & 565 & 26,466 & 168 & 5,890 & 44 & 4,011 & 62 \\
10-30 km & 49,247 & 259 & 5,947 & 33 & 567 & <10 & 407 & <10 \\
>30 km & 38,467 & 214 & 2,720 & 15 & 27 & <10 & 35 & <10 \\
\hline
Total & $\approx$233,857 & $\approx$1,239 & 213,231 & 1,419 & 50,023 & 578 & 39,674 & 705 \\
\end{tabular}
\caption{Number of student pairs (N) and student pairs co-infected within 14 days of each other (N\_inf) as a function of distance and background. G1: Pairs of students who did not attend the same primary (in 2020) or secondary schools (2021). G2--4: Attended the same primary school in 2020 and (G1) did not attend the same secondary school in 2021; (G2) attended the same secondary school but different program; (G3) attended the same secondary school and the same program. Note that results can only be exported from the secure computer of CBS in groups of at least 10
individuals.}\label{t:summary_pairs}
\end{table}

\begin{table}[ht!]
\begin{tabular}{r|llllll}
Group & \multicolumn{2}{l}{Siblings} & \multicolumn{2}{l}{Parent-child} & \multicolumn{2}{l}{Co-parents} \\
 & N & N\_inf & N & N\_inf & N & N\_inf \\
\midrule
Same house & 888,288 & 205,326 & 1,447,899 & 328,507 & 807,156 & 330,502 \\
<0.3 km & 111,504 & 8,000 & 88,401 & 10,067 & 9,196 & 796 \\
0.3-1 km & 290,834 & 16,638 & 225,625 & 20,560 & 32,278 & 1,848 \\
1-3 km & 491,886 & 24,418 & 347,810 & 27,298 & 58,324 & 3,026 \\
3-10 km & 732,272 & 28,004 & 426,277 & 27,962 & 73,072 & 3,188 \\
10-30 km & 631,268 & 18,324 & 313,655 & 15,626 & 53,754 & 1,690 \\
>30 km & 939,994 & 22,542 & 419,501 & 16,659 & 52,748 & 1,008 \\ \hline
Total & 4,086,078 & 323,254 & 3,269,180 & 446,679 & 1,086,532 & 342,058 \\
\end{tabular}
\caption{Number of family pairs (N) and family pairs co-infected within 14 days of each other (N\_inf) as a function of distance and type of family tie.}
\label{t:summary_pairs_family}
\end{table}

\subsection{Statistical Analysis of School and Municipality Heterogeneity}\label{s:m:regression}

In our second analysis we focused on the factors driving the transmission dynamics between students. We conducted a multilevel regression analysis, where we modeled temporally associated infections between student pairs using logistic regressions, with parameters estimated via maximum likelihood estimation.

Multilevel models are capable of accurately estimating regression parameters in situations where data is hierarchically structured, and thus violating the assumption of independence of observations. They furthermore allow to attribute variance to the respective levels in the data structure.

We constructed models accounting for a three-level structure: individuals, schools, and municipalities (Dutch: \textit{gemeenten}). This enabled us to identify the extent to which the school context contributes to temporally associated infection events, separating it from influences at the individual and municipal level. 

We added several explanatory variables at each level to explain variability in temporally associated infection probabilities. At the individual level, we added the \textit{distance} between student pairs as a predictor of an associated infection. Due to the skewed distribution of the variable, and to aid convergence in parameter estimation, we took its natural logarithm, centered, and z-scaled it (i.e., subtracted the mean and divided by the standard deviation to normalize the values). The school-level predictors were the number of students indicating the \textit{size} of the school, the median \textit{income} of the school's 4-letter postcode area, and the school's \textit{denomination} (if any). School size and income were centered and z-scaled for the same reasons as the distance variable.

A first model served as the baseline, decomposing the variance at the different levels by including random intercepts for schools and municipalities, but not including any predictors:

\begin{equation}
         y_{isg} = \gamma_{000} + v_{0m} + u_{0sm} + e_{ism},
    \label{eq:mlm1}
    \end{equation}

where $i$ denotes an individual, $s$ a school, and $m$ a municipality. In the equation, $\gamma_{000}$ is the overall intercept and $v_{0m}, u_{0sm}, e_{ism}$ represent the error terms at the municipality, school, and individual level, respectively.

We then estimated a second model including the random intercepts introduced above as well as the predictors at the individual and school level:

\begin{equation}
         y_{ism} = \gamma_{000} +  \gamma_{p00} X_{pism} + \gamma_{0q0} Z_{qsm} + v_{0m} + u_{0sm} + e_{ism},
         \label{eq:mlm2}
\end{equation}

where $X_{pism}$ corresponds to the ($p=1$) predictors at the individual level: the logged distance between students. $Z_{qsm}$ represents the ($q=3$) predictors at the school level: school size, median income, and school denomination.

Finally, we included random slopes ($u_{psm}$) at the school level for the distance between student pairs ($X_{pism}$):
\begin{equation}
         y_{ism} = \gamma_{000} +  \gamma_{p00} X_{pism} + \gamma_{0q0} Z_{qsm} + u_{psm} X_{pism} + v_{0m} + u_{0sm} + e_{ism}.
        \label{eq:mlm3}
\end{equation}

Significance of predictors was assessed using Wald tests at $\alpha=0.05$. Significant differences between the models were determined by likelihood-ratio tests. Explained variance at different levels of the models was calculated according to the method proposed by \textcite{mckelvey_statistical_1975}, which relates the systemic variance of the model introduced by the predictor variables to the residual variance at all levels. Model coefficients and variances were furthermore rescaled following \textcite[pp.125]{hox_multilevel_2017} to enable comparison of explained variance across models (see also \textcite[pp.121-125]{hox_multilevel_2017} for more clarification on variance calculation and rescaling procedures in multilevel models for dichotomous outcomes.).

Instead of running the three models for all the different groups of student pairs introduced in Section \ref{s:m:matching}, we restricted this part of the study to students who attended primary education together (in the same class year) in 2021. Furthermore, the analyses were based on a 5-percent sample of all schools in the data, with an inclusion probability proportional to school size. This was done in order make the models computationally feasible, while still including a substantial number of different schools from various areas. The sampling yielded a dataset of 2,509,927 observations representing student pairs, grouped in 312 schools, and 174 municipalities. While there is a large class imbalance, with 0.1\% of the student pairs temporally co-infected, we follow the advice of recent research of not correcting for it \parencite{van2022harm}. Coefficient estimates are stable for high class imbalance as long as there are enough observations in the minority class and corrections tend to miscalibrate the models \parencite{van2022harm}.

The statistical models were run using the \textit{lme4} library in R. The Python and R code documenting the performed steps of all data processing and analysis procedures is available at \url{https://github.com/jgarciab/covid_schools/}.

%% RESULTS
\section{Results and Discussion}\label{s:results}

\subsection{Shared School and Classroom Environments}

We first analyzed the role of schools in the transmission of SARS-CoV-2 by focusing on students transitioning from primary school to secondary school (see Methods Section~\ref{s:m:matching}), which allowed us to better separate school from non-school social interactions such as community transmission.

\begin{figure}[h!]
    \centering
    \includegraphics[width=\textwidth]{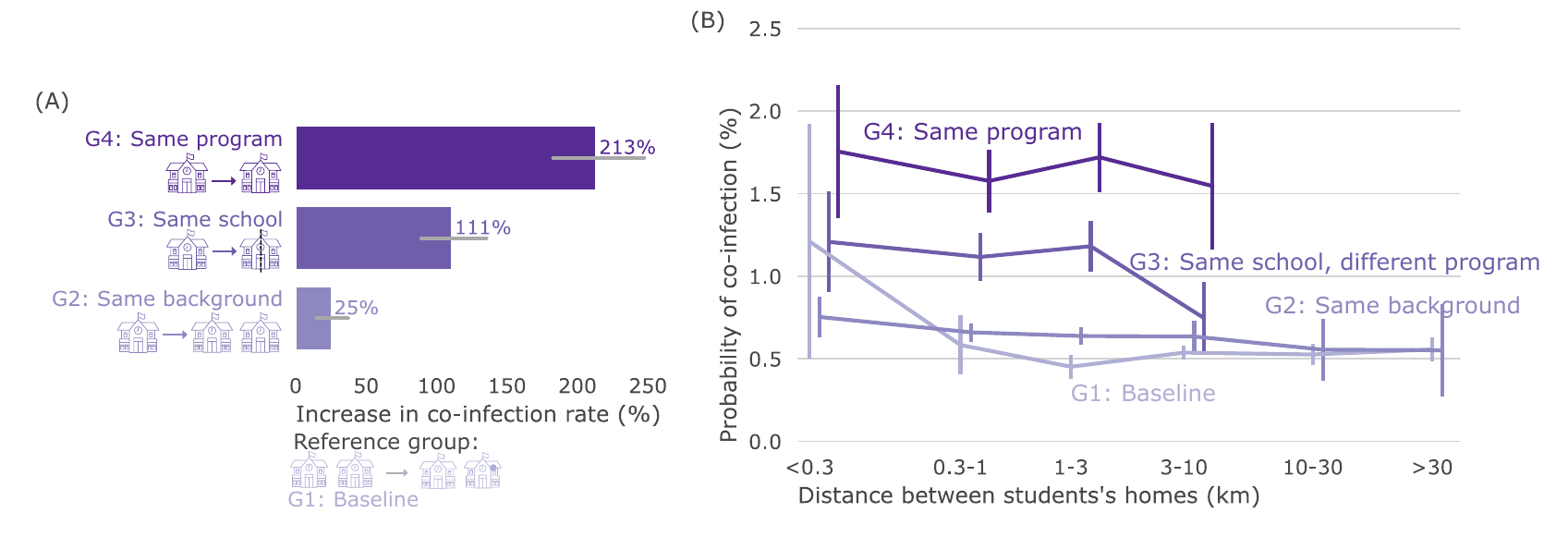}
    \caption{\small \textbf{Attending the same school increases the probability of temporally associated infection.} (A) Increase in the probability of temporally associated infection, compared to the baseline (G1), for student pairs in the same program track (G4), same school but different program track (G3), and different school but same background (G2). Error bars indicate 95\% confidence intervals. (B) Probability of temporally associated infection for the four different groups of student pairs, as a function of the distance between the student's homes. Note the logarithmic horizontal axis and that the $<0.3km$ distance bin excludes individuals living in the same household. Groups have slightly different horizontal offsets to avoid overlapping error bars.}
    \label{fig:school}
\end{figure}

We found that the probability of temporally associated infection was 0.52\% (95\% confidence interval (CI): 0.49--0.56\%) for the baseline group, 1.11\% (CI: 1.01--1.20\%) for students in group 3, and 1.65\% (CI: 1.52--1.77\%) for students in group 4. Compared with the baseline group (group 1), attending the same primary school and the same program track in secondary school (group 4) increased the probability of associated infections significantly by 213\% (CI: 183--247\%, see Fig.~\ref{fig:school}A). Attending secondary school in a different program track (group 3)---thus not sharing a classroom so frequently---increased the probability of associated infections significantly by 111\% (CI: 89--135\%, see Fig.~\ref{fig:school}A). 

Students who attended primary school together but not secondary school together (group 2) had only a slightly higher probability of temporally associated infections (CI: 0.62--0.69\%) compared to the baseline (0.52\%). This small difference indicates that social ties inherited from primary school had little impact on this probability. Moreover, for all groups of students, the distance between students' houses had only a minor effect on the probability of associated infection (Fig.~\ref{fig:school}B). %This result suggests that interventions targeting out-of-school activities may have limited impact on reducing SARS-CoV-2 transmission in educational settings.

\subsection{Shared Household and Family Contexts}
After assessing the increase in the probability of temporally associated infections for students attending the same school, we examined how this probability compares to the probability for individuals of the same family.

\begin{figure}[h!]
    \centering
    \includegraphics[width=\textwidth]{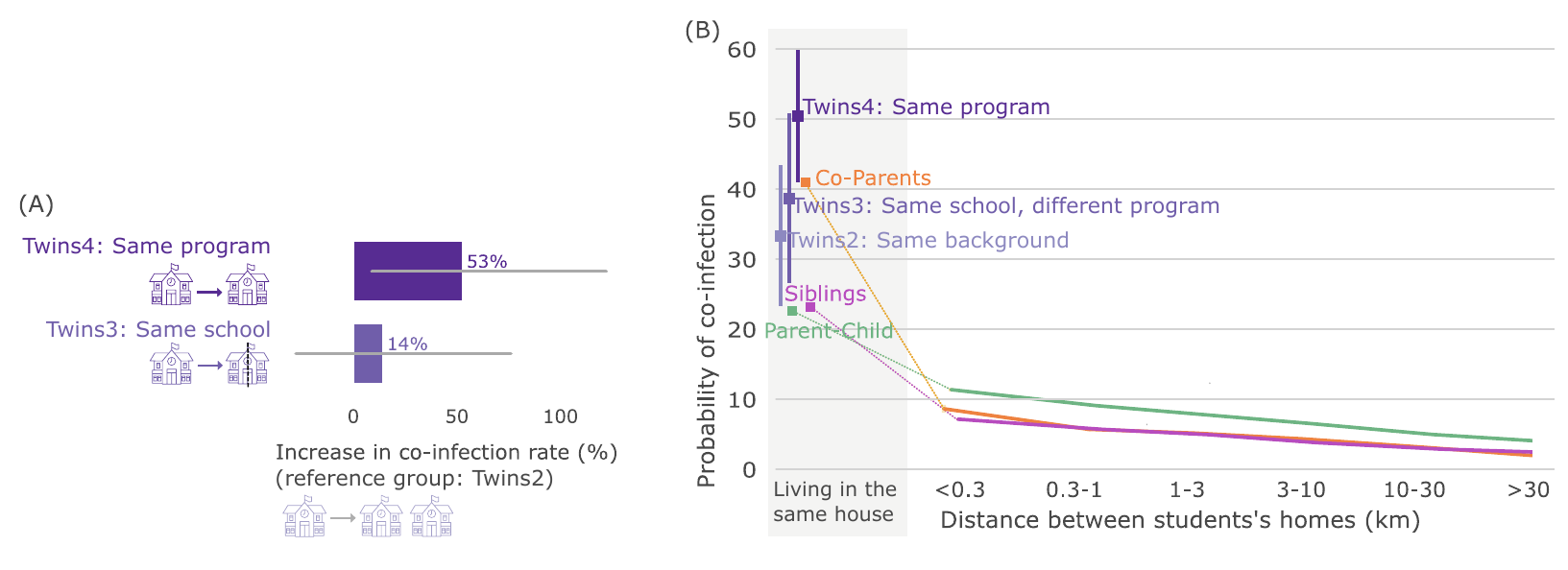}
    \caption{\small \textbf{Comparison of the probability of temporally infections in school and family networks.} (A) Increase in temporally associated infection rate, compared to the baseline (Twins2: siblings attending a different school but having attended the same primary school), for twins in the same program (Twin4), and same school but different program track (Twins3). Error bars indicate 95\% confidence intervals. (B) Probability of temporally associated infection as a function of the distance between the individual's homes for twin pairs (purple), sibling pairs (magenta), parent-child pairs (green) and co-parents (orange). Groups have slightly different horizontal offsets to avoid overlapping error bars. Note the logarithmic horizontal axis. The grayed area correspond to pairs living in the same household and a break in the logarithmic axis.}
    \label{fig:family}
\end{figure}

For twins living in the same household, attending the same school modestly increased the probability of temporally associated infection: 33\% (28 out of 84 pairs, CI: 28–84\%) for twins attending different secondary schools, compared to 39\% (24 out of 62 pairs, CI: 24–62\%) for twins attending the same school and different track, and 50\% (54 out of 107 pairs, CI: 54–107\%) for twins attending the same school and program track (Fig.~\ref{fig:family}A). Living in the same household results in a large increase in the probability of temporally associated infection (Fig.~\ref{fig:family}B). The estimated probabilities ranged from 23\% for sibling pairs and parent-child pairs to 50\% (CI: 41--60\%) for twins attending the same program track in secondary school. This increased risk is presumably due to prolonged exposure if there is an active case. The probabilities for family pairs are much larger than the estimated 1.6\% for students in group 4 (attending the same year group in primary and secondary school). This finding aligns with current scientific knowledge highlighting the key role of household transmission in the spread of SARS-CoV-2 (see for example \cite{vanierselEmpiricalEvidenceTransmission2023, cmmidcovid-19workinggroupImplicationsSchoolhouseholdNetwork2021c, cordery2022transmission, thompson2021staff}).

Among individuals who do not share a household, family relationships are highly predictive of the probability of temporally associated infections (Fig.~\ref{fig:family}B). The probability of temporally associated infections for parent-child, co-parents and siblings living in different but nearby households is 7--12\%. This probability decreases with distance (Fig.~\ref{fig:family}B), as social interactions between family members are more likely when they live close together.

\subsection{School and municipality heterogeneity in the probability of temporally associated infections}

We finally investigated the determinants of temporally associated infections through a series of multilevel regression models (Table~\ref{tab:mlm}), which explicitly attribute variance to different levels of observational units (see Section \ref{s:m:regression}). The presented results of Model 1---including random intercepts for schools and municipalities but no predictors---serves as a baseline to assess the variance explained by the predictors. The majority of the total variance in this three-level structure manifested at the individual level with $\frac{3.29}{3.29+1.93+0.25} = 0.60$ (i.e., 60\% of the variance is due to individual-level differences). The school-level variance made up a share of 0.35, which left 0.05 to the municipality level. 

\begin{table}[ht!]
\centering
\footnotesize
\begin{tabular}{@{}llll@{}}
\toprule
DV: Temporally associated infections (within 14 days)                       & Model 1   & Model 2   & Model 3   \\ \midrule
Intercept                                            & -6.77 (0.09)*** & -6.35 (0.28)*** & -6.42 (0.25)*** \\
log(Distance)                                        &           & -0.12 (0.02)***  & -0.30 (0.05)*** \\
School size                                          &           & -0.14 (0.18)     & -0.13 (0.18)    \\
Median income                                        &           & -0.08 (0.11)     & -0.06 (0.11)    \\
\textit{School denomination:}                        &           &                  &    \\
- Anthroposophic/Waldorf [ASF]                       &           & -0.90 (0.58)     & -0.82 (0.71)     \\
- Evangelical [EVA]                                  &           & 1.00 (1.10)      & 1.24 (1.10)      \\
- Reformed (liberated) [GEV]                         &           & -0.76 (0.80)     & -0.67 (0.77)    \\
- Inter-confessional [IC]                            &           & 0.12 (1.01)      & 0.19 (0.98)      \\
- Islamic [ISL]                                      &           & -2.81 (1.27)*   & -2.66 (0.83)**  \\
- Non-denominational [OPB]                           &           & -0.49 (0.29)     & -0.46 (0.28)     \\
- Protestant [PC]                                    &           & -0.74 (0.32)*   & -0.71 (0.32)*   \\
- Reformed Christian [REF]                           &           & -0.29 (0.57)    & -0.23 (0.55)    \\
- Roman Catholic [RK]                                &           & -0.29 (0.29)    & -0.25 (0.28)     \\
- Cooperative Catholic/Protestant [SPR]              &           & 0.07 (0.9)      & 0.16 (0.88)     \\ \midrule
AIC                                                  & 71811.80  & 71772.55  & 71576.41  \\
BIC                                                  & 71850.01  & 71976.32  & 71805.65  \\
Log   Likelihood                                     & -35902.90 & -35870.28 & -35770.20 \\
Number   observations                                & 2,509,927   & 2,509,927   & 2,509,927   \\
Number   schools (within municipality)               & 312       & 312       & 312       \\
Number   municipalities                              & 174       & 174       & 174       \\ \midrule
\textit{Decomposition of variance:} &           &           &           \\ 
- Individual (Random Intercept)                              & 3.29      & 3.19      & 3.19      \\
- School (Random Intercept)                                 & 1.93      & 1.81      & 1.81      \\
- Municipality (Random Intercept)                           & 0.25      & 0.13      & 0.13      \\
- Predictors                                         &           & 0.16      & 0.24      \\
- School:distance (Random Slope)                            &           &           & 0.18      \\
Covariance between school intercept and slope                &           &           & 0.04      \\ \bottomrule
\end{tabular}
\caption{\small Results of logistic multi-level regression models. Statistically significant coefficients are marked as *** $p<0.001$, ** $p<0.01$, * $p<0.05$. Standard errors are displayed next the coefficients in parenthesis. The school denominations were translated from Dutch. Codes of the original variable from the CBS dataset \textit{INSCHRWPOTAB} are displayed in square brackets. The reference category of denomination is \textit{Specialized non-denominational education [ABZ]} (e.g., Montessori). Numerical variables were centered and z-scaled.}
\label{tab:mlm}

\end{table}

We then included predictors indicating the residential distance of student pairs, the size of the school they attended, the median income of the postcode area of the school, and the school's denomination (Model 2). The random effects remained the same as in Model 1---i.e., including random intercepts for schools and municipalities. 

The distance between student pairs' homes was found to significantly decrease the temporally associated infection probability. This is in line with the results of the preceding probability analysis, identifying student pairs living in the same household to be facing the highest risk of temporally associated infection. We also found significant decreases in associated infection probabilities in Islamic and Protestant schools as compared to the reference category. While Islamic schools are only a small minority of all schools, Protestant, together with public schools and Catholic schools, are one of the largest denominations. However, our analysis does not distinguish if the results are driven by a lower spread of virus or by a lower propensity to test. School size and median income in the postcode area did not show a significant association with temporally associated infections. 

In total, the predictors were able to explain 3\% of the overall variance. This can be calculated by the share of variance in the linear predictor compared to the total variance of Model 2, or comparing the total variances of Model 2 and Model 1. Looking at variance reduction at all levels individually, variance decreased by 3 percent at the individual level, 6.2 percent at the school level, and, most notably, by 48 percent at the municipality level. While the predictors explained a large share of variability at the municipality level, the variability was very low to begin with (5\% of the total). A likelihood-ratio test confirmed a significant improvement in model fit ($\chi^2(13) = 65.24, p < 0.001$) of Model 2 over Model 1.

Finally, in Model 3, we included random slopes of student-pair distance at the school level---i.e., the effect of distance was allowed to vary by school. A term for intercept-slope covariance was also included, modeling the strength of the distance effect depending on the average probability by school. These parameters substantially increased model fit, meaning, the association of distance between the students residence and temporally associated infection probability was indeed dependent on the specific school. This is also indicated by the significant result of the likelihood-ratio test comparing Model 3 to Model 2 ($\chi^2(2) = 200.16, p < 0.001$).

To conclude, the variance decomposition of the multi-level models showed that the vast majority of variability in the data results from differences at the individual level (60\%) and the school level (35\%). While we could find significant effects of student distance and school denomination on temporally associated infection probability, these effects could explain only 3 percent of the overall variance. Possible omitted factors driving differences in this probability could be families' attitudes towards COVID-19, or prevention measures implemented at the school level.

%% CONCLUSION
\section{Conclusion}\label{s:conclusion}
In this paper we investigate the impact of schools and families in the temporal association of SARS-CoV-2 infections among students during the period from June 2020 to September 2021. This is possible by integrating population-scale networks and PCR test result data using registry data from Statistics Netherlands.

Our results show that living together at home is the most significant factor correlated with two individuals testing  positive within a 14-day period, underscoring the importance of household transmission in the spread of the virus. Both social ties inherited from primary school and geographical distance were found to have little effect on the probability of both students testing positive within a 14-day period. This suggests that either social ties with classmates in primary school are weakened after students move to secondary school, or that COVID restriction strategies targeting non-school social networks were highly effective. Future studies could estimate the effect of school and non-school restrictions on students contacts and SARS-CoV-2 transmission.

In contrast with the low impact of social ties inherited from primary school, shared school and classroom environments were found to significantly increase the likelihood that both students would test positive within a 14-day period. Although the likelihood of temporally associated infections in schools was low, it should be noted that even small increases in the transmission rate in schools can lead to larger outbreaks, since the transmission rate is linearly related to the reproduction number \parencite{millerNoteDerivationEpidemic2012} and a large proportion of children's contacts are expected to occur in schools. The observed increase in temporally associated infections from 0.6 to 1.6\% may lead to reproduction numbers above one infectee per infected when schools reopen \parencite{mundayEstimatingImpactReopening2021}. These insights into the transmission dynamics of SARS-CoV-2 within Dutch families and educational institutions can inform future use of network models and provide insights for possible interventions, such as school closures. 

The analysis presented in this paper has a limitation that open up fruitful additional avenues for future research. Governments around the world introduced several interventions to reduce the transmissions, including (partially) closing schools, workplaces, and wearing masks in confined spaces. Due to the wide variety of measures at school, workplaces and public spaces \parencite{rozhnova2021model}, we did not examine the role of school closures. Using data from Statistics Netherlands and a similar methodological approach, further research could explore the effectiveness of various interventions. Furthermore, a similar approach merging infection results data and population scale data could be used to understand the effects of school and family connections for other diseases.

% \subsection*{CRediT author statement}
% Conceptualization and writing: All authors. Methodology and Formal analysis: Javier Garcia-Bernardo, Christine Hedde-von Westernhagen. Funding acquisition: Tom Emery

%% ACKNOWLEDGMENTS
\subsection*{Acknowledgments} \label{ack}

This work was financed by the Ministry of Health, Welfare and Sport (Ministerie van Volksgezondheid, Welzijn en Sport: H16-4068-27762) and facilitated by ODISSEI (\url{https://odissei-data.nl/}), the Dutch National Infrastructure for Social Research. 

% \subsection*{Author contributions} 
% Conceptualization: all authors; methodology, data curation and analysis: J.G.B., C.H.vW;.;  writing—original draft preparation: J.G.B., C.H.vW; writing— review and editing: all authors; visualization: J.G.B; supervision: J.G.B., T.E., A.J.vH. project administration: J.G.B and T.E..; funding acquisition: T.E.

\subsection*{Ethical statement} 
Data is collected by Statistics Netherlands (CBS) and the National Institute for Public Health and the Environment (RIVM), and made available to researchers for well-defined projects and statistical analysis. Researchers need to be pre-approved before accessing the data, and all data is pseudoanomyzied, and available in a secure research environment. The data is safeguarded under the stringent privacy regulations set by the Statistics Netherlands Act (``Wet op het Centraal bureau voor de statistiek'') and the European Union's General Data Protection Regulation, guaranteeing that individual personal information is not revealed during the analysis. All methods were carried out in accordance with relevant guidelines and regulations.

\subsection*{Data availability} 
%rephrased from christine's thesis
Administrative data from Statistics Netherlands (CBS) is available for statistical and scientific research under specific conditions. For more details about the data and its usage, as well as access and usage regulations, we refer you to \textcite{bokanyiAnatomyPopulationscaleSocial2022, vanderlaanWholePopulationNetwork2022}. For inquiries regarding micro-data, please contact microdata@cbs.nl.

Aggregated data exported from CBS and all analysis script can be found at \url{https://github.com/jgarciab/covid_schools/}. For further needs contact the corresponding author.

\subsection*{Declaration of Competing Interest}
The authors declare no competing interests.

\subsection*{Declaration of generative AI and AI-assisted technologies in the writing process}
During the preparation of this work the author(s) used \url{https://www.deepl.com/write} for copyediting and to improve readability. After using this tool/service, the author(s) reviewed and edited the content as needed and take(s) full responsibility for the content of the publication.

%% BIBLIOGRAPHY
\printbibliography

% % Fill out appendix:
\newpage
\appendix
\begin{refsection}
\section{Supplementary Information}\label{a:appendix1}
\subsection{Detailed explanation of datasets and variables used}
All datasets at individual level available at Statistics Netherlands (CBS) are linkable to each other through a unique identifier (the combination of the variables \textit{RINPERSOON} and \textit{RINPERSOONS}, which together identify individuals in the data).

\subsubsection{COVID-19 PCR tests \textit(CORONIT/CoronIT\_GGD\_testdata\_20210921) dataset}
The COVID-19 dataset includes test results from PCR tests conducted outside of hospitals. Key variables include the individual’s persistent identifier (\textit{RINPERSOON}), test date (\textit{DatumMonsterafname}), and test result (\textit{Testuitslag}). Reinfections were uncommon during the study period, so we retained only the first infection date for each individual.

\subsubsection{Person Network  (\textit{PN/PersNw2018\_v1.0}) dataset}
At the time of the study, the Person Network dataset compiled by CBS was only available for 2018. We used this dataset to extract family relations. This dataset includes  the identifier (\textit{RINPERSOON(S)}) of the two individuals and the type of connection (e.g. \textit{103} for full-siblings, \textit{102} for co-parents, \textit{104} for parent-child relationships). 

\subsubsection{Registrations in primary school (\textit{Onderwijs-INSCHRWPOTAB}) dataset}
To assess if students attended primary education (ages 4 to 12) together, we used the dataset of student registrations in primary schools (\textit{Onderwijs-INSCHRWPOTAB}), keeping the school category (\textit{WPOTYPEPO}) ``BO'' or ``Basisonderwijs'' (the variable label was changed by CBS, but there was no change in the school systems). This only excludes children that require extra help in special schools---such as blind or deaf students---and whose infection patterns are likely to be different. Each school is associated to a school denomination, \textit{WPODENOMINATIE}. The school denomination denotes the type of school and is correlated with attitudes towards COVID-19. In the Netherlands, parents have the right to choose schools that match their values. A majority of schools are Christian (either Protestant, Catholic, Evangelic or Reformist), while around one third are public schools \parencite{engzell2021learning}. Other denominations include Islamic schools and Anthroposophic.

Students attended primary school together if they attended the same educational site (variable \textit{WPOBRIN\_crypt}) at the same branch (variable \textit{WPOBRINVEST}, which allow us to distinguish between schools with different locations) and at the same year of education (variable \textit{WPOLEERJAAR}). 

\subsubsection{Registrations in secondary school (\textit{Onderwijs-ONDERWIJSINSCHRTAB}) dataset}
To assess if students attended secondary education (students aged 12+) together, we used the dataset \textit{Onderwijs-ONDERWIJSINSCHRTAB}. This dataset contains all registrations in secondary schools. Similarly to the approach in primary schools, we kept only the school category (variable \textit{TYPEONDERWIJS}) ``VO'' and excluded registrations in special schools. This dataset provides only registration of students by date of registration (variable \textit{AANVINSCHR}) and deregistration (variable \textit{EINDINSCHR}), not by year of study. Since we focus only on students in the first year of secondary school, registration date and year of study are expected to be almost identical---i.e, we assume students did not repeat year, which is highly unlikely for primary school. We kept students registering after August (and before September of the next year) in the same study year. We removed 1,884 students that were not registered for at least half a year in any school. This could happen for example if the student attended three or more schools in a year. 

Students attended secondary school together if they attended the same educational site (variable \textit{BRIN\_crypt}), school branch (variable \textit{VOBRINVEST}, which allow us to distinguish between schools with different locations), year of study derived from the date of registration (\textit{AANVINSCHR}) as explained above, and program track of education (variable \textit{OPLNR}), which allows us to distinguish students in the (pre-)vocational, general, applied, and scientific tracks. Students in different tracks attend different classrooms.

\printbibliography[heading=subbibliography,title={Bibliography}]
\end{refsection}

\end{document}